# A Domain Specific Ontology Based Semantic Web Search Engine

**Debajyoti Mukhopadhyay[1, 2]    Aritra Banik[1]   Sreemoyee Mukherjee[1]   Jhilik Bhattacharya[1]**

[1] Web Intelligence & Distributed Computing Research Lab, Techno India,
West Bengal University of Technology
EM 4/1, Salt Lake Sector V, Calcutta 700091, India
debajyoti.mukhopadhyay@gmail.com

**Young-Chon Kim[2]**

[2] Advanced Communications & Networks Lab, Division of Electronics & Information Engineering
Chonbuk National University
561-756 Jeonju, Republic of Korea

## ABSTRACT

*Since its emergence in the 1990s the World Wide Web (WWW) has rapidly evolved into a huge mine of global information and it is growing in size everyday. The presence of huge amount of resources on the Web thus poses a serious problem of accurate search. This is mainly because today's Web is a human-readable Web where information cannot be easily processed by machine. Highly sophisticated, efficient keyword based search engines that have evolved today have not been able to bridge this gap. So comes up the concept of the Semantic Web which is envisioned by Tim Berners-Lee as the Web of machine interpretable information to make a machine processable form for expressing information. Based on the semantic Web technologies we present in this paper the design methodology and development of a semantic Web search engine which provides exact search results for a domain specific search. This search engine is developed for an agricultural Website which hosts agricultural information about the state of West Bengal.*

## 1. INTRODUCTION

A **search engine** is a document retrieval system designed to help find information stored in a computer system, such as on the World Wide Web, inside a corporate or proprietary network, or in a personal computer. The search engine allows one to ask for content meeting specific criteria (typically those containing a given word or phrase) and retrieves a list of items that match those criteria. Regardless of the underlying architecture, users specify keywords that match words in huge search engine databases, producing a ranked list of URLs and snippets of Web-pages in which the keywords matched. Although such technologies are mostly used, users are still often faced with the daunting task of sifting through multiple pages of results, many of which are irrelevant.

Surveys indicate that almost 25% of Web searchers are unable to find useful results in the first set of URLs that are returned [6].

Tim Berners-Lee, the inventor of the World Wide Web, defines the *Semantic Web* as "The Web of data with meaning in the sense that a computer program can learn enough about what the data means [in order] to process it" [1].

Rather than a Web filled only with human-interpretable information, Berners-Lee's vision includes an extended Web that incorporates *machine interpretable* information, enabling machines to process the volumes of available information, acting on behalf of their human counterparts [7].

In this paper, we discuss the basic idea of the semantic Web and describe a design and development methodology for a domain specific semantic Web search engine based on ontology matching which not only overcomes the problem of knowledge overhead but also supports complex queries. Further, it is able to produce exact answers that in one hand satisfy user queries and on the other hand are self-explanatory and understandable by end users.

## 2. SEMANTIC WEB SEARCH ENGINE

### 2.1 The working of a regular search engine
For most internet users, a search engine is the starting point of finding desired information in the Web. The most common form of text search used by the majority of popular search engines on the Web is **keyword based search** that is**,** they do their text query and retrieval using keywords. The working of any regular search engine may be summarized as follows:

- Search engine searches its enormous database for the keyword - entered by the user (after pressing the search button.)
- Every engine has its own collection system to fill its database.
- Indexing system is used to organize the database - permits faster searching
- Returns a list of hit -includes relevant (as well as irrelevant) pages

This keyword based search technique gives rise to several problems listed as follows:

- The Web is growing much faster than any present-technology search engine can possibly index. In 2006, some users found major search-engines became slower to index new Web-pages.
- Keyword searches have a tough time distinguishing between words that are spelled the same way, but mean something different. This often results in hits that are completely irrelevant to the query.
- Some search engines also have trouble with stemming, i.e., if the word "big,"

is entered, should it return a hit on the word, "bigger?" What about singular and plural words? What about verb tenses that differ from the word someone entered by only an "s," or an "ed"?

- Search engines also cannot return hits on keywords that mean the same, but are not actually entered in the query. A query on heart disease would not return a document that used the word "cardiac" instead of "heart."
- Users are returned thousands to millions of Web pages in return of their queries, of which majority prove to be irrelevant to the query submitted and is impossible for any user to go through.

In view of the above mentioned problems, come up the concept of semantic Web and semantic Web search engines.

### 2.2 Semantic Web and Semantic Search Engine
"The Semantic Web is the representation of data on the World Wide Web. It is a collaborative effort led by W3C with participation from a large number of researchers and industrial partners. It is based on the Resource Description Framework (RDF), which integrates a variety of applications using XML for syntax and URIs for naming." – W3C Semantic Web. The Semantic Web is a framework that allows publishing, sharing, and reusing data and knowledge on the Web and across applications, enterprises, and community boundaries [4]. Currently, the Semantic Web, consisting of Semantic Web documents typically encoded in the languages RDF and OWL, is essentially a Web universe parallel to the Web of HTML documents [5]. Knowledge encoded in Semantic Web languages such as RDF differs from both the largely unstructured free text found on most Web pages and the highly structured information found in databases. Such semi-structured information requires using a combination of techniques for effective indexing and retrieval. RDF and the Web Ontology Language (OWL) which are ontology based procedures or representing knowledge on the Web, introduce aspects beyond those used in ordinary XML, allowing users to define terms

(for example, classes and properties), express relationships among them, and assert constraints and axioms that hold for well-formed data. An application of the emerging Semantic Web is a Semantic Web search engine which searches the Semantic Web documents against a user query for accurate results. Our work uses RDF encoded Semantic Web documents which are searched in response to a user query for exact results.

## 3. ONTOLOGY

An ontology is an explicit specification of some topic. For our purposes, it is a formal and declarative representation which includes the vocabulary (or names) for referring to the terms in that subject area and the logical statements that describe what the terms are, how they are related to each other, and how they can or cannot be related to each other. Therefore, Ontology provides a vocabulary for representing and communicating knowledge about some topic and a set of relationships that hold among the terms in that vocabulary [2] [3].

**Why develop an Ontology?**

- To enable a machine to use the knowledge in some application.
- To enable multiple machines to share their knowledge.
- To help yourself understand some area of knowledge better.
- To help other people understand some area of knowledge.
- To help people reach a consensus in their understanding of some area of knowledge.

In our project we used Resource Description Framework or RDF to represent knowledge. For example, if we need to describe a subject in terms of its classes and their relationships using RDF, we are creating an Ontology.

As our project deals with the crops domain, the designed ontology is shown in Figure 1. In Figure 2, the general information ontology is depicted. A relational diagram is shown in Figure 3 to depict some classes, instances, and relations among them in the crops domain.

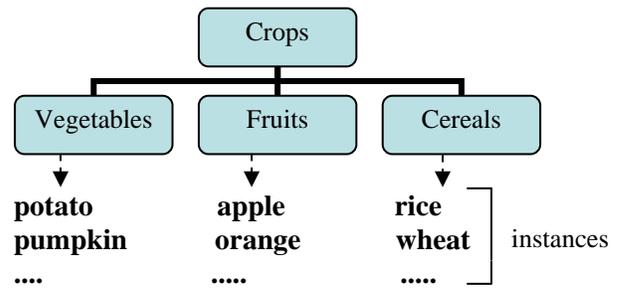

**Figure 1.** The Crops ontology: Crops is the superclass of its subclasses vegetables, fruits, and cereals

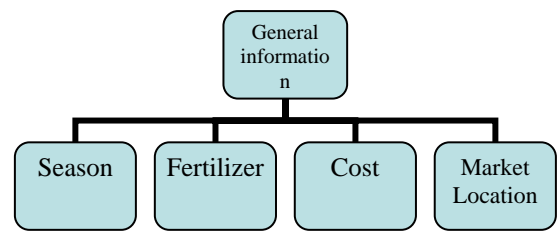

**Figure2.** The general information ontology: General information is the superclass of subclasses: season, fertlizer, cost, market location

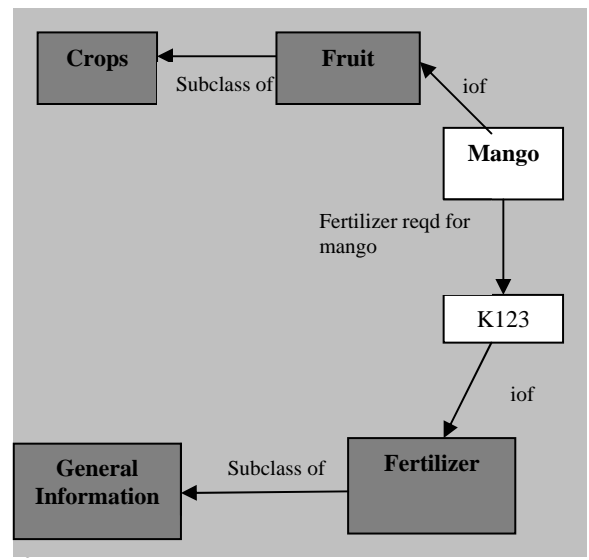

**Figure 3.** Some classes, instances, and relations among them in the crops domain; shaded boxes are used to indicate *classes* and non-shaded for *instances*; iof indicates *instance of*

## 4. OUR APPROACH

Based on the ontology as described above, we have first developed the relevant RDF pages.

### 4.1. What is RDF (Resource Description Framework)

- Resource Description Framework (RDF) is a framework for describing and interchanging metadata (data describing the Web resources) [8] [11].
- RDF provides machine understandable semantics for metadata.

This leads to,

- better precision in resource discovery than full text search
- assisting applications as schemas evolve
- interoperability of metadata

Components of RDF data model: [9]

- Directed labeled graphs
- Model elements:

**Resource:** Everything described by RDF expressions is called a resource [10]. Every resource has a URI and it may be an entire Web page or a part of a Web page.

**Property:** "A property is a specific aspect, characteristic, attribute, or relation used to

Describe a resource" – W3C, Resource Description Framework (RDF) Model and Syntax Specification [10]. Note that a property is also a resource since it can have its own properties.

**Statements:** A statement combines a resource, a property and a value. These three individual parts are known as the "subject," "predicate" and "object".

For example, "Potato requires the soil type KR256" is an RDF statement having the following parts:

Subject (Resource)       Potato
Predicate (Property)     soilreqd
Object (value)           KR256

This can be represented as an RDF graph as shown in Figure 4.

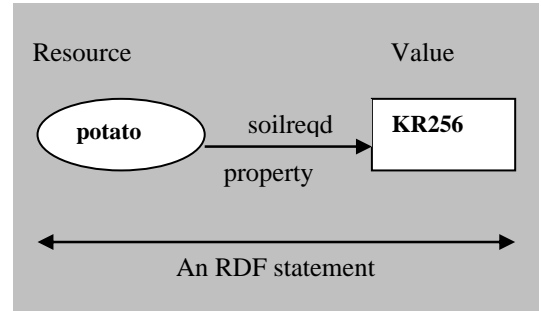

**Figure 4.** An RDF Statement

```
<?xml version="1.0"?>
<vegetable rdf:ID="potato"
xmlns:rdf="http://www.w3.org/199
9/02/22-rdf-syntax-ns#"
xmlns="http://www.westbengal.org
/crops#">
<soilreq>KR256</soilreq>
</vegetable>
```

This RDF code describes the resource "potato" which in an instance of the "vegetable" class. "KR256" is the value of the property "soilreq." Similar documents have been created for all the classes and instances of each class as specified in the ontology. A separate RDF code describes the class subclass relationships.

**RDF Schema:** RDF Schema is a simple data-typing model for RDF [12] so that we can describe groups of related resources and the relationships among these resources [11]. For example, we can say "potato" is a type of "vegetable" and "vegetable" is a subclass of "crops." The purpose of RDF schema is to express classes and their (subclass) relationships as well as to define properties and associate them with classes. The benefit of an RDF Schema is that it facilitates inferencing on the data, and enhanced searching.

Resources can be divided into "**classes**" which is composed of instances. A class itself is also a resource which is usually identified by **RDF URI References** and can be described by **RDF properties**. We often use the prefix

**"rdfs:"** to indicate the term is RDF Schema term. **"rdfs:resource"** is the root class of everything in RDF Schema. **"rdf:type"** is an instance of rdf:Property (class of RDF properties), and it means that a resource is an instance of a class. The property **rdfs:subClassOf** is an instance of rdf:Property that is used to state that a class is a Subclass of the other. The following RDF code shows that "Crops" is the super-class and "Vegetable" and "Fruits" are subclasses of "Crops."

```
<?xml version="1.0"?>
<rdf:RDF
xmlns:rdf="http://www.w3.org/1999/02/22-rdf-syntax-ns#"

xmlns:rdfs="http://www.w3.org/2000/01/rdf-schema#"
        xml:base="http://www.westbengal.org/crops#">
<rdfs:Class rdf:ID="Vegetable">
  <rdfs:subClassOf
rdf:resource="#Crops"/>
</rdfs:Class>
<rdfs:Class rdf:ID="Fruits">
  <rdfs:subClassOf
rdf:resource="#Crops"/>
</rdfs:Class>
    ...
    ...
</rdf:RDF>
```

Another important RDF technique that we have utilized in our design methodology is the specification of the RDF domain and range.

**RDF Domain and Range: Domain** is used to indicate the classes that a property will be used with. One may specify zero, one, or multiple rdfs:domain properties [10][11].
**Range** is used to indicate the type of values that a property will contain. One may specify zero, one, or multiple rdfs:range properties.

The following RDF code snippet shows the usage of domain and range. It describes an RDF property "seasonreqd" whose domain is the vegetable class i.e. it can be used with this class and its range is the "season" class i.e. it can take values of the type "season" only.

```
<rdf:Property
rdf:ID="seasonreqd">
    <rdfs:domain
rdf:resource="#Vegetable"/>
    <rdfs:range
rdf:resource="#season"/>
</rdf:Property>
```

Thus, using RDF schema a class is defined separately and its relations to other classes are indicated, again properties are defined separately, associated with the respective classes, having their domains and ranges specified.

**4.2. The Design of the Semantic Web Search Engine**
The basic components of the ontology based semantic Web search engine is shown in Figure5.

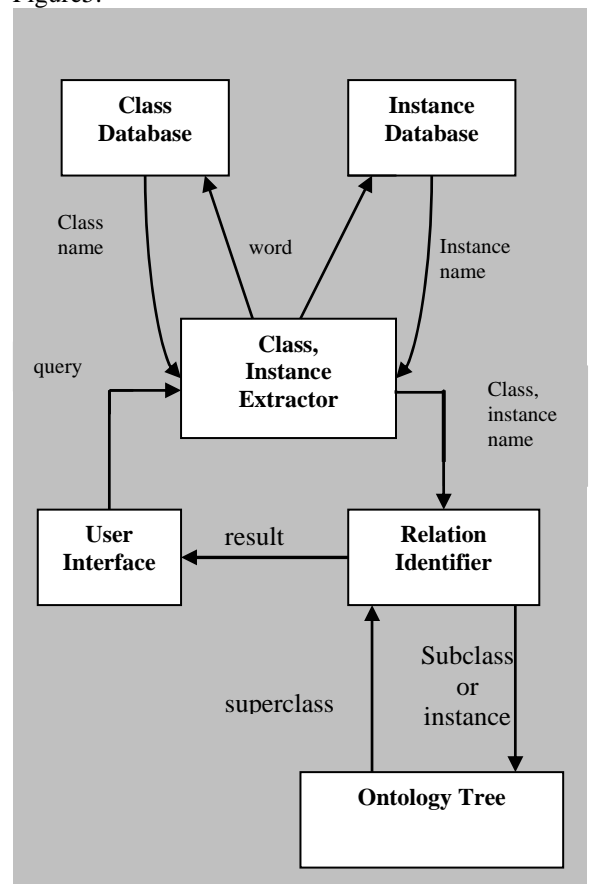

**Figure 5.** Basic Components of the semantic Web search engine

As shown in the Figure 5, the main parts of our design of the search engine are:

- A Class, instance extractor
- A class database
- An instance database
- An Ontology tree
- A Relation identifier

An user interface is designed where the user enters his query and the results are returned back to him. As the search engine works for a specified domain, it directly retrieves the results rather than the URLs (as returned by a conventional search engine) and displays them. The different modules of the search engine are discussed below.

**The Class Instance Extractor:** This module is responsible for the extraction of class or instance names present , if any , in the query entered by the user i.e. it takes as input the user query and returns as output the class or instance name or both, if present in the query, which match with those defined in the RDF codes. For e.g., from a query "season required for mango," it will extract the words "season" and "mango" as they correspond to the class "season" and the instance "mango" defined in the RDF codes. It does so with the help of the class database and the instance database by checking each word of the user query with the class and instance names stored in these databases.

**The Class Database:** This module is a simple database which contains all the names of the classes used in the RDF codes and their corresponding synonyms. It is implemented as a two column table where one column stands for the class names and the other its corresponding synonym. A class name may have one or more synonyms. It checks each word supplied to it by the class instance extractor if it matches with the words stored. If yes, it returns that word back to the class instance extractor.

**The Instance Database:** The function and implementation of this module is similar to the class database except that it stores the instance names as defined in the RDF codes.

**The Ontology Tree:** This module is nothing but the entire ontology, stored in the form of RDF codes as described previously. It is the storage of the class subclass instance relationships as RDF codes to facilitate the working of the relation identifier, which uses this module to find out the class of an instance or the superclass of a subclass etc.

**The Relation Identifier:** This module performs the most important task of identifying relationships between the class and instance extracted from a query by the class instance extractor. It tries to relate the extracted class and instance by a predefined property i.e., a property defined in the RDF codes. Consider the query "season required for mango." The relation identifier will first look up all the property definitions to find a property whose domain is mango and range is season. As mango is an instance and not a class, it will not be able find any such property. It will then move up one level upwards in the ontology tree to find the class of the instance "mango" and will try to find a property whose domain is fruits and range is season. It will find the property season_req, predefined in the RDF code for mango, which relates mango and season. A simplified diagrammatic representation of this mapping is shown in the Figure 6.

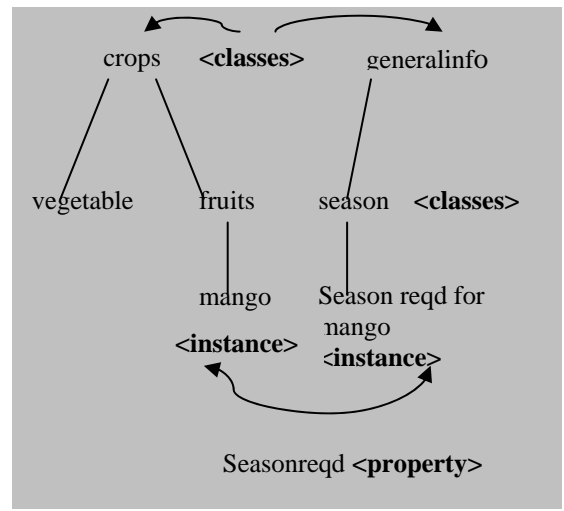

**Figure 6:** Mapping from instance to Class

On being able to successfully relate the class and the instance the relation identifier will also find the value of this property from the RDF codes and will return that value as the final result.

In any Natural Language Processing model, a parser is an obvious part. It helps to identify the semantic correctness of the sentence or query. But in case of an ontology based design, the parsing rules are already present in the form of RDF codes. The property itself identifies the mapping as it includes its domain and range as attributes in the RDF codes where they are defined. If a direct relation in the form of range and domain is not present, we have to look further in the ontology tree in order to identify the relation or mapping. Thus our ontology based semantic Web search engine also works for indirect queries , or queries which do not contain a class  or instance name directly.

Suppose the query is: "K123 required for which crops?" K123 stands for a fertilizer. In this case the relation identifier gets from the class instance extractor the words "K123" and "crops." Therefore it first checks whether there is any property whose domain is K123 and range is crops. If not present, it goes one level upward in the ontology tree and checks if there exists any property whose domain is fertilizer and range is crops. Here the property is defined as *fertilizerreqd*. As the property name is known, it searches every subclass of crops and their instances in order to identify the instances for which the value of  the property *fertilizerreqd* is K123 .

On the other hand, if a query "crops required for which K123" is given, it replies back an error message as there is no property present whose domain is fertilizers and range is crops.

Thus our semantic Web search engine gives accurate, meaningful, exact results of a user query.

## 5. EXPERIMENTAL RESULTS

In our prototype the domain specific ontology based search engine is developed as an extendable J2EE application where each of the modules as specified above have been developed using java. An user interface is also developed as a part of the crops Website where the user enters his query for search. Some screenshots of the user interface containing the search query and its results are shown in Figure 7 and Figure 8.

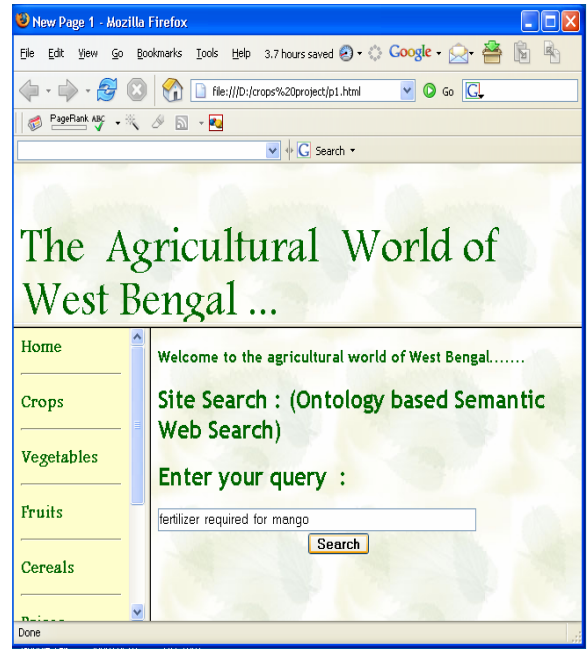

**Figure 7.** Given query is "Fertilizer required for mango"

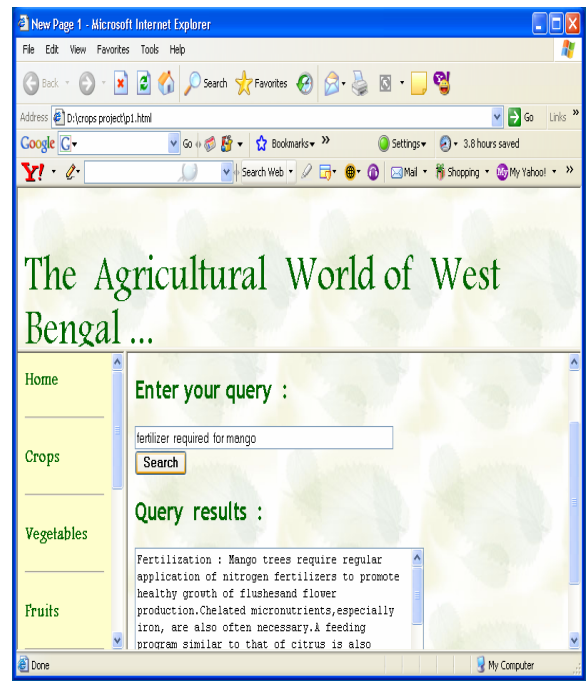

**Figure 8.** Results for the query "Fertilizers required for mango"

## 5.1. Performance Analysis:

It can be shown that even in the worst case the performance of the semantic Web search engine is much better than a regular search engine with increasing number of pages. It works faster than a regular search engine when the world wide Web is considered to be divided into several domains of similar ontologies (Figure 9). For a general semantic Web search engine, the task remains only to search domain-wise and hence find the resultant page from the Web.

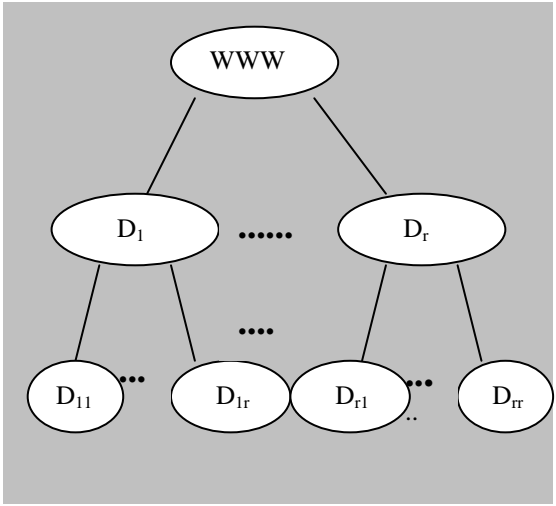

**Figure 9.** The World Wide Web divided into domains

Let us consider the world wide Web to be divided into say 'r' domains ($D_1$ to $D_r$) containing Web sites or pages of similar ontologies. Let each domain be further divided into 'r' domains. Node $D_1$ divided into domains $D_{11}$ to $D_{1r}$. Let the height of this tree be 'h' and the total number of elements in this tree be 'n' which is equal to the total number of Web-pages present in the Web.

Now, from the above tree,

$$n = 1 + r + r*r + r*r*r + \ldots\ldots + r^h$$

$$n = \frac{r^h - 1}{r - 1}$$

$$or, h = \frac{\log(n(r-1)+1)}{\log r}$$

$$or, h = \log_r (n(r-1)+1)$$

To derive this equation, we have considered number of subdomains for each and every domain is 'r'. But in reality, the number of subdomains for each domain may not be constant. The number of subdomains explicitly depends on the design schema. Here we have concentrated for the average case analysis assuming r as the average number of sub-domains for each domain

**Best Case:** In this case, the query follows a straight path as a domain path, i.e. the query starts from $D_1$ then goes to $D_{11}$, $D_{111}$...so on. Therefore, the complexity of search is in the order of h, which is equal to the order of $\log_r (n(r-1)+1)$.

**Worst Case:** In this case, the search topic requires at most 'r' searches per level to determine the domain of the query. Therefore the number of searching required is r(h-1) which is equal to $r*(\log_r (n(r-1)+1)-1)$.

## 5.2 Graph Plotting:

Here, by the term performance we have referred to the number of searches needed for a successful query processing. Other factors like storage capacity or query parsing time can be ignored for a domain containing huge number of pages such as the World Wide Web.

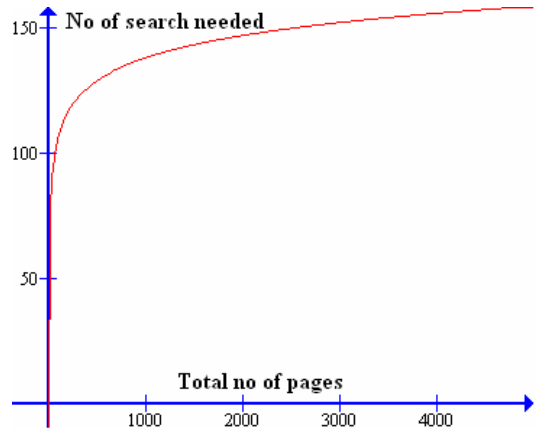

**Figure 10.** Performance graph of the semantic Web search engine for r=50.

For a regular search engine, to identify a single query it requires 'n' number of searches where n is the total number of pages present. In Figure 10, we have shown the total number of searches needed for different number of pages. For instance, from the above graph it is clear for a domain containing 1000 pages, search needed for a semantic web search engine for a query is about 138. On the other hand for regular search engine we must search for all 1000 pages. See Figure 11.

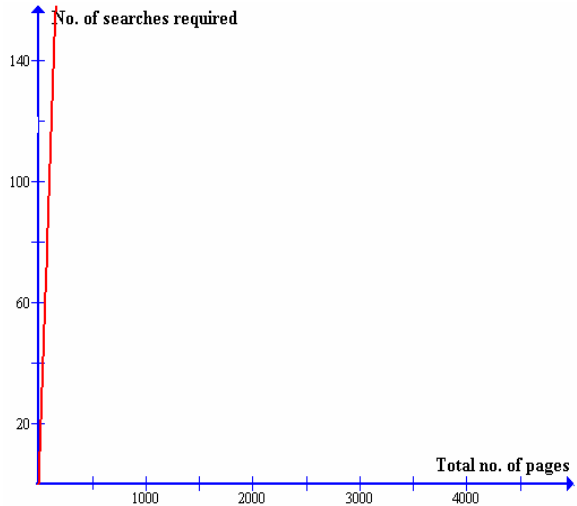

**Figure 11.** Performance graph of a regular keyword based search engine

From the above graphs it is clear although for a small domain the search needed for both the cases is nearly equal. But as we increase the number of pages, the performance of a semantic search engine becomes better. Therefore for a large domain like the WWW, performance of a semantic web search engine is extraordinarily better than a regular search engine.

## 6. CONCLUSIONS AND FUTURE WORK

Though new improved keyword-based technologies for searching the WWW are evolving constantly, the growth rate of these improvements is likely to be slight. Problems of imprecise and irrelevant results will continue to hinder Web searchers, especially with the continued expansion of the Web. Search engines based on a new concept as the semantic Web technology, are effectively able to handle the above mentioned problems. A domain specific ontology based semantic search engine as ours is advantageous in several ways.

Firstly, our approach has been able to successfully eliminate the problem of irrelevant results which is one of the main problems encountered by the users of a regular search engine. By using the mapping technique between instances and classes, the search engine effectively fetches the exact information.

Secondly, by producing exact information as the result, the search engine eliminates the need to go through numerous results as in case of a regular search engine.

Thirdly, the number of searches and time required by the semantic search engine is less than that of a regular search engine.

Lastly, our design although based on the crops domain, is highly scalable and can be easily adopted by other enterprises as their site search tool. This would only require the enterprise to feed in the relevant RDF codes based on the ontology of the domain. As a result the page containing the site search (ontology based semantic search) would be automatically generated.